\documentclass{iaus}
\usepackage{graphicx}

\title[Mass loss rates inferred from mid-IR color excesses of LMC and SMC 
O stars]
{IR mass loss rates of LMC and SMC O stars}

\author[D.\ Massa et al.]   
{D. Massa$^1$, A. Fullerton$^1$, D. Lennon$^1$
 \and R.K. Prinja$^2$}

\affiliation{$^1$STScI, Baltimore, MD, USA  \\[\affilskip]
$^2$Dept. of Astronomy, UCL, London, UK}

\pubyear{2010}
\volume{272} 
\pagerange{1--2}
\jname{Active OB stars: structure, mass loss and critical limits}
\editors{C. Neiner, G. Wade, G. Meynet \& G. Peters, eds.}
\begin{document}

\maketitle

\begin{abstract}
We use a combination of VJHK and {\it Spitzer} [3.6], [5.8] and [8.0] 
photometry, to determine IR excesses in a sample of LMC and SMC O stars. 
This sample is ideal for determining excesses because: 1) the distances to 
the stars, and hence their luminosities, are well-determined, and; 2) the 
very small line of sight reddenings minimize the uncertainties introduced 
by extinction corrections.  We find IR excesses much larger than expected 
from Vink et al.\ (2001) mass loss rates.  This is in contrast to previous 
wind line analyses for many of the LMC stars which suggest mass loss rates 
much less than the Vink et al.\ predictions.  Together, these results 
indicate that the winds of the LMC and SMC O stars are strongly structured 
(clumped).
\end{abstract}

\firstsection 
\section{Introduction}
The effects of structure on wind diagnostics is now widely accepted.  
It causes H$\alpha$ and radio and IR continuum measurements ($n_e^2$\ 
diagnostics) to {\em over}-estimate mass loss rates, $\dot{M}$.  
Structure, and its accompanying porosity, can also cause wind line 
diagnostics to {\em under}-estimate $\dot{M}$ (Prinja \& Massa 2010).  
We examine this effect in lightly reddened Magellanic Cloud O stars by 
deriving $\dot{M}$s from {\it Spitzer} IR excesses, and comparing the 
results to theoretical expectations from Vink et al.\ (2001).  If clumping 
is significant, we expect the IR excesses to over-estimate the mass loss 
rates.  Furthermore, comparing the LMC and SMC results allows us to examine 
metallicity effects on clumping.

\section{Extinction Corrections} 
To determine IR excesses, the continua must be corrected for reddening, and 
this requires the assumption that a portion of the continuum is free of 
excess and has a known slope, typically $(B-V)$.  Applying an extinction 
curve with an inappropriate $A(V)$ can introduce enormous errors in the 
inferred IR excess.  In this regard, LMC and SMC O stars have a substantial 
advantage compared to Galactic O stars.

\section{The Model}
We use a generalization of the Lamers \& Waters (1984) model which 
agrees with results from the Puls et al.\ (1996) Fastwind model when 
similar parameters are used.  We ignore the effects of disks or 
non-standard, slowly accelerating velocity laws at this time.   For 
the stellar parameters the following were adopted: For the LMC we 
used the Martins et al.\ (2002) calibrations for the stellar parameters, 
except for luminosities, where a distance modulus of 18.52~mag was used.  
We assumed $Z(\rm{LMC})/Z_\odot = 0.6$.  For the SMC we used the Massey 
et al.\ (2009) Spectral Type~$\rightarrow T_{eff}$ calibration, a $DM = 
18.91$~mag to determine $\log L(\rm{SMC})/L_\odot$, and the Letherer et 
al.\ (2010) $Z(\rm{SMC})/Z_\odot = 0.2$ grid to determine masses.  When 
measured terminal velocities, $v_\infty$, were not available, we used the 
Vink et al.\ (2001) formulae relating escape velocity and $v_\infty$.  
TLUSTY (Lanz \& Hubeny 2003) models of the appropriate temperatures, 
surface gravities and metallicities were used for the bare photospheres, 
and an $R(V) = 3.1$ extinction curve was assumed throughout (Fitzpatrick 
\& Massa 2009).  

\section{Results}
All of the IR mass loss rates (see, Fig.~1) are larger than expected and, 
as is well known, LMC mass loss rates are larger than SMC rates for similar 
stellar parameters.  Furthermore, the relative disagreement between the IR, 
$\dot{M}$(IR), and Vink et al.\ (2001) mass loss rates, $\dot{M}$(Vink), is 
similar for the LMC and the SMC.

\begin{figure}[h]
\begin{center}
\includegraphics[width=2.5in]{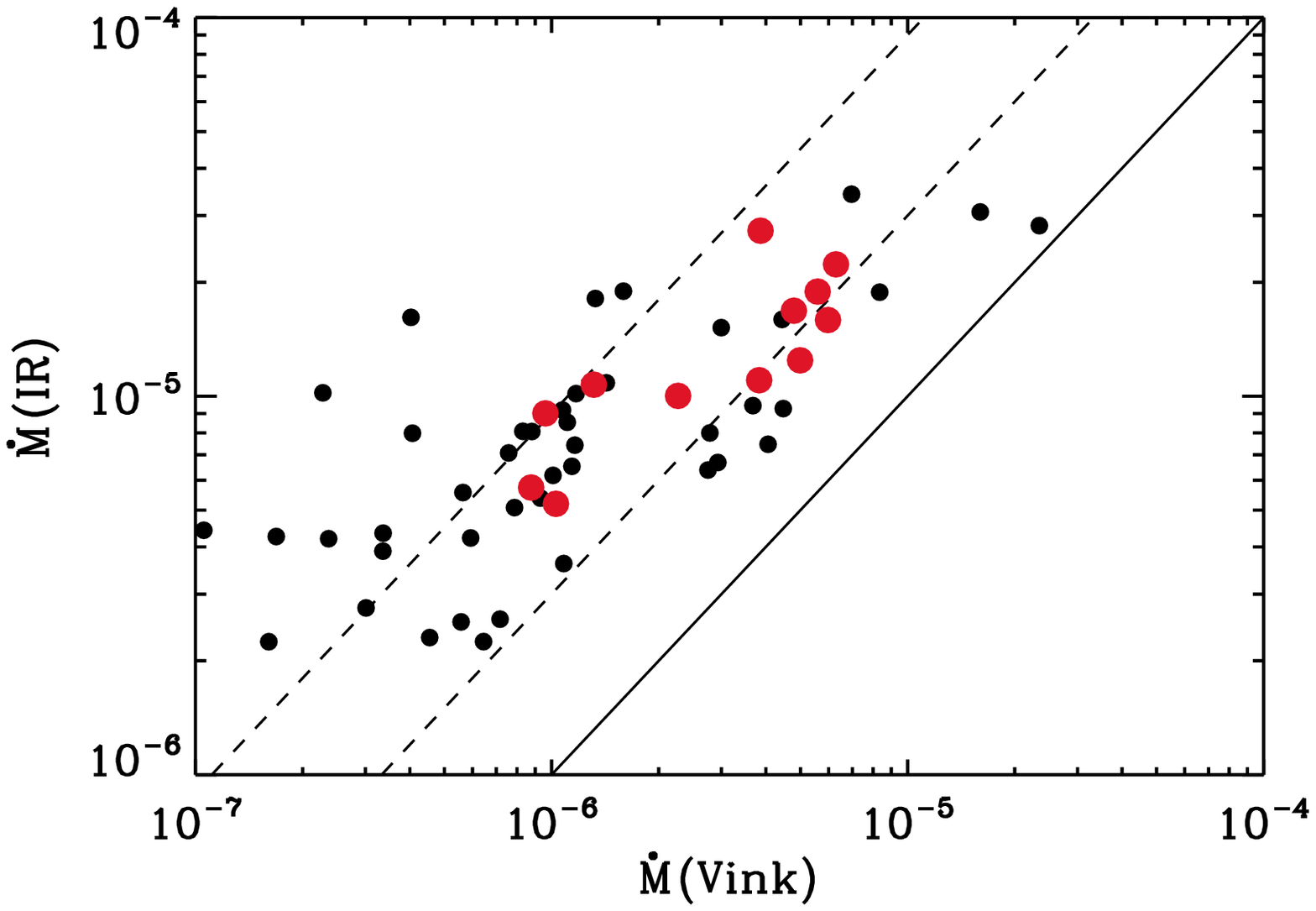}\hfill 
\includegraphics[width=2.5in]{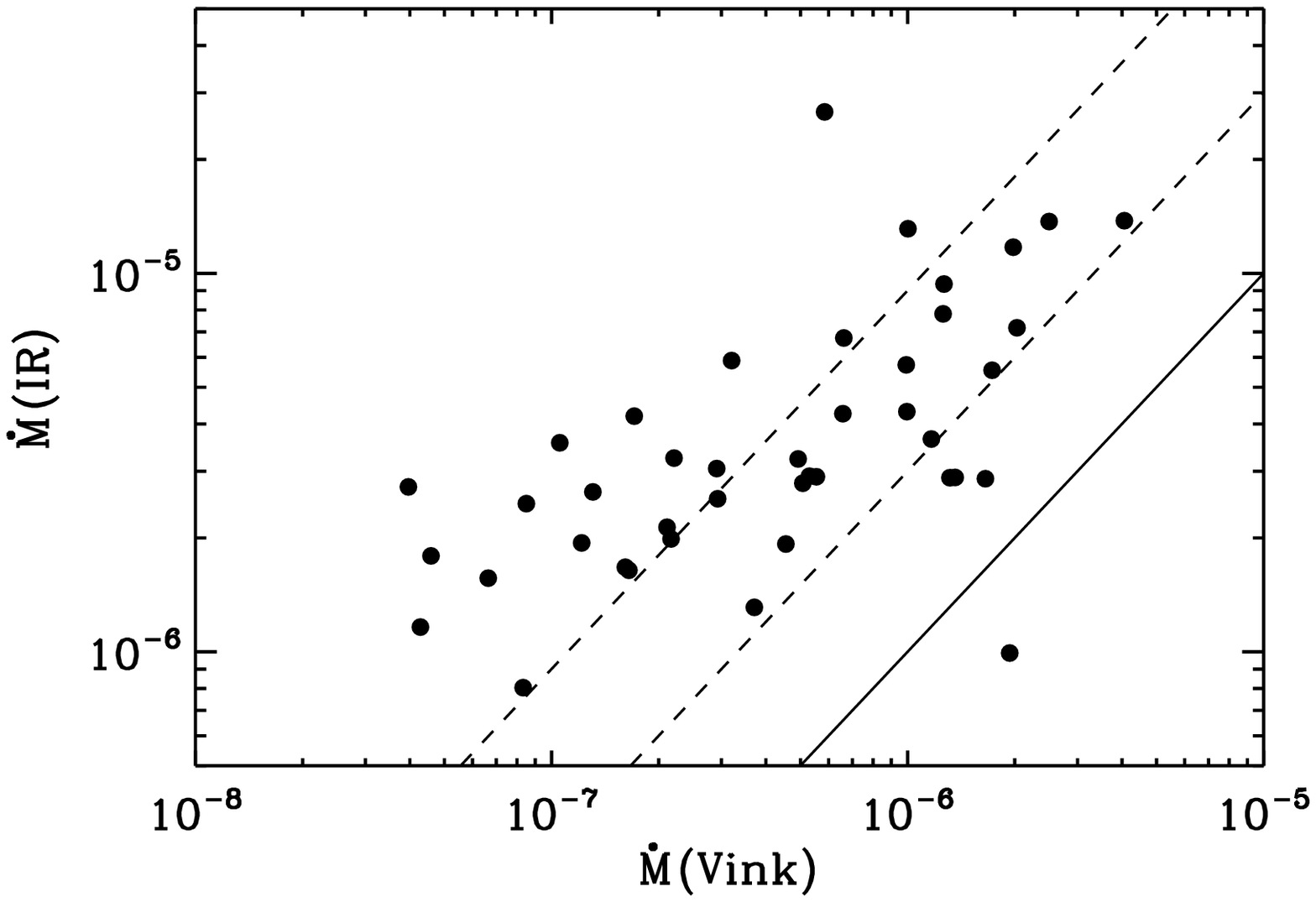} 
\caption{Mass loss rates from {\it Spitzer} excesses, $\dot{M}$(IR), as a 
function of theoretical expectations, $\dot{M}$(Vink), for the LMC (left) 
and SMC (right).  For the LMC, P~{\sc V} results, infer $\dot{M}$s much 
smaller than $\dot{M}$(Vink).  The red points depict stars which will be 
observed by {\it HST}.  The solid line indicates $\dot{M}$(IR$) = 
\dot{M}$(Vink) and the dashed lines give $\dot{M}$(IR$) = 3$ and 9$\times 
\dot{M}$(Vink). } \label{fig1}
\end{center}
\end{figure}

\section{Conclusions}
\noindent $\bullet$ IR excesses for LMC and SMC O stars are larger than 
those expected from theory by factors of 3 -- 10. \\
$\bullet$ For LMC stars, the mass loss rates are vastly larger than those 
expected from UV wind lines (Massa et al.\ 2003, Fullerton et al.\ 2006).\\
$\bullet$ The relative disagreement between theory and observation is 
similar for both galaxies. \\
$\bullet$ If, as expected, clumping causes the disagreement, then its 
effect appears to be weakly dependent on metallicity.

\end{document}